# Exploring the Thermostability of CRISPR–Cas12b using Molecular Dynamics Simulations

Yinhao Jia[1], Katelynn Horvath[2], Santosh R. Rananaware[1], Piyush K. Jain[1,3,4], Janani Sampath[1,*]

[1] Department of Chemical Engineering, University of Florida, Gainesville, FL, USA

[2] Department of Chemical and Biomolecular Engineering, University of Connecticut, Storrs, CT, USA

[3] Department of Molecular Genetics and Microbiology, University of Florida, Gainesville, FL, USA

[4] Health Cancer Center, University of Florida, Gainesville, FL, USA

* jsampath@ufl.edu

## Abstract

CRISPR (clustered regularly interspaced short palindromic repeat)- based diagnostics are at the forefront of rapid detection platforms of infectious diseases. The integration of reverse transcription-loop-mediated isothermal amplification (RT-LAMP) with CRISPR-Cas protein systems has led to the creation of advanced one-pot assays. The sensitivity of these assays has been bolstered by the utilization of a thermophilic Cas12 protein, BrCas12b, and its engineered variant, which exhibits enhanced thermal stability and allows for broader operation temperatures of the assay. Here, we perform all-atom molecular dynamics (MD) simulations on wild-type and mutant BrCas12b to reveal the mechanism of stabilization conferred by the mutation. High-temperature simulations reveal a small structural change along with greater flexibility in the PAM-interacting domain of the mutant BrCas12b, with marginal structural and flexibility changes in the other mutated domains. Comparative essential dynamics analysis between the wild-type and mutant BrCas12b at both ambient and elevated temperatures provides insights into the stabilizing effects of the mutations. Our findings not only offer a comprehensive insight into the dynamic alterations induced by mutations but reveal important motions in BrCas12b, important for the rational design of diagnostic and therapeutic platforms of Cas12 proteins.

**Introduction**

CRISPR (clustered regularly interspaced short palindromic repeats) – Cas (CRISPR-associated protein) systems have emerged as revolutionary tools in the field of molecular biology. This system, originally a part of bacterial immune defenses, has been adapted to perform a myriad of tasks in genetic engineering, due to its precision and versatility.[1,2] One of the latest advances in CRISPR/Cas involves using Cas proteins as components in molecular diagnostics, with applications ranging from infectious disease detection to genetic disorder diagnosis and cancer detection.[3,4] Cas9 and dCas9 have been used to establish diagnostic systems for the detection of Single Nucleotide Polymorphisms (SNPs) and featured nucleic acid sequences in pathogenic strains.[5,6] The discovery of indiscriminate cleavage of other non-specific single-stranded RNAs upon target binding, also referred to as trans-cleavage or collateral cleavage activity of CRISPR-Cas13a (C2c2) in 2016 brought a new era of CRISPR/Cas-based biosensing.[7] Cas13a was utilized to establish a comprehensive diagnostic platform, SHERLOCK, which was further developed for rapid diagnosis of infectious diseases.[8–10]

In addition to Cas13, Cas12a (Cpf1) was found to have collateral cleavage activity on other single-stranded DNAs. Specifically, upon binding to the intended nucleotide target, a non-target or collateral cleavage is triggered in the Cas protein which is distinct from specific cleavage activity.[11,12] Collateral activity can be utilized to cleave DNA reporter, thereby magnifying the fluorescence signal and the sensitivity of the detection platform; this Cas12a-based nucleic acid detection platform, DETECTR, was reported by Chen et al.[11] At the same time, Li et al. discovered a similar collateral cleavage effect in Cas12a and developed the nucleic acid sensing assay HOLMES.[13] While the specificity and the simplicity of these nucleic acid detection assays are advantages, the requirement of separate steps to amplify the sample before the detection to increase sensitivity poses significant challenges, including possible carryover contamination, longer processing time, complexity in handling, often an increase in the overall cost of the assay. To address this, the integration of Cas proteins with a pre-amplification step, such as reverse transcription loop-mediated isothermal amplification (RT-LAMP), has shown promise, as the integration allows for a sensitive nucleic acid detection method.[14,15] However, the high-temperature nature of RT-LAMP, typically conducted at 65°C, denatures most native Cas proteins, hindering proper integration to create a simple one-step process.

With the utilization of the Cas12b (C2c1), a thermophilic RNA-guided endonuclease from Class II type V-B CRISPR/Cas, Wang and others showed that with the same cleavage activity, pairing with LAMP (loop-mediated isothermal amplification) results in a one-step assay for convenient target detection.[16,17] More recent efforts including the usage of AapCas12b, which is derived from *Alicyclobacillus acidiphilus*, and BrCas12b, which derived from *Brevibacillys sp.*, as part of the one-step diagnostic assay for the detection of nucleic acids including SARS-CoV-2. Based on SHERLOCK, Joung et al. integrated RT-LAMP with AapCas12b, thereby achieving a one-step detection that exceeds the sensitivity of CDC RT-qPCR.[18] Nguyen et al. simplified the detection process and reduced the carryover contamination by a one-pot assay developed using the newly identified thermostable Cas12b ortholog BrCas12b (with a melting temperature of 62 °C).[19] Further optimization of the single-pot assay was achieved by the combination of *de novo* design through point mutations and experimental validation, resulting in a thermostable BrCas12b with a melting temperature of 67 °C (referred to as eBrCas12b) which does not degrade at the high operating temperatures (65 °C) of RT-LAMP.[20] While these modifications have shown promising results, the underlying mechanisms of how these mutations contribute to increased thermal stability is not fully understood, nor do we know its influence on the overall dynamics of the protein, which could affect its cleavage process.

Molecular dynamics (MD) simulations is a powerful tool to gain insights into the structure and dynamics of proteins, including the effects of mutations at the atomic level.[21,22] MD simulations have been implemented previously to understand fundamental activation mechanism of CRISPR-Cas9 systems, revealing high conformational flexibility of the HNH domain and providing insight into the dynamic behavior of Cas9. Palermo et al. performed a series of MD simulations on CRISPR-Cas9 systems to explore the conformational plasticity of the Cas9 protein and its interactions with nucleic acid during DNA binding.[23] These studies revealed that the Cas9 protein undergoes significant conformational changes during the process of nucleic acid binding and processing. Meanwhile, the 'closure' of the protein relies on the highly coupled and specific motion between different domains. Saha et al. investigated the structural dynamics of different Cas12a orthologs and found that DNA binding induces changes in Cas12a conformational dynamics, which leads to the activation of REC2 and Nuc domains that enable the cleavage of nucleic acids, particularly the large amplitude motion of Nuc that allows the protein to open and close. The dynamics of REC2 and Nuc domain are highly coupled, and REC2 can act as a regulator of Nuc, which is similar to HNH in Cas9 system.[24] While CRISPR-Cas9 has been extensively analyzed using MD simulations,[25] there remains a significant gap in understanding other Cas proteins, such as Cas12. Considering the unique collateral cleavage mechanism of Cas12 and the difference in cleavage mechanism between Cas12b and Cas12a,[26,27] a thorough analysis of Cas12b is needed for rationally designing variants of the protein. Particularly, MD simulations can provide insights into BrCas12b and eBrCas12b, shedding light on how mutation improves the stability of BrCas12b at high temperatures and how the new point mutations alter the essential dynamics of

BrCas12b. This information is important to design superior CRISPR analogs for future diagnostic and therapeutic applications.

We focus on the BrCas12b ortholog of Cas12b and employ all-atom MD simulations to probe its structure and dynamics before and after mutation. By conducting a series of high-temperature unfolding simulations, we reveal the unfolding process of wild type and mutant BrCas12b, highlighting the structure and flexibility change of specific domains containing mutations, which substantially improve the stability and activity at high temperatures. Additionally, we explore the essential dynamics of wild-type and mutated-type BrCas12b at both ambient and elevated temperatures together with cross-correlation analysis, examining the internal movements within the protein and interactions between its domains, revealing the change of essential dynamics as well as decreased correlation between domains upon mutation. These results contribute to a fundamental understanding of structural dynamics of BrCas12b and the influence of mutations on the protein function.

## Methods

### System Setup

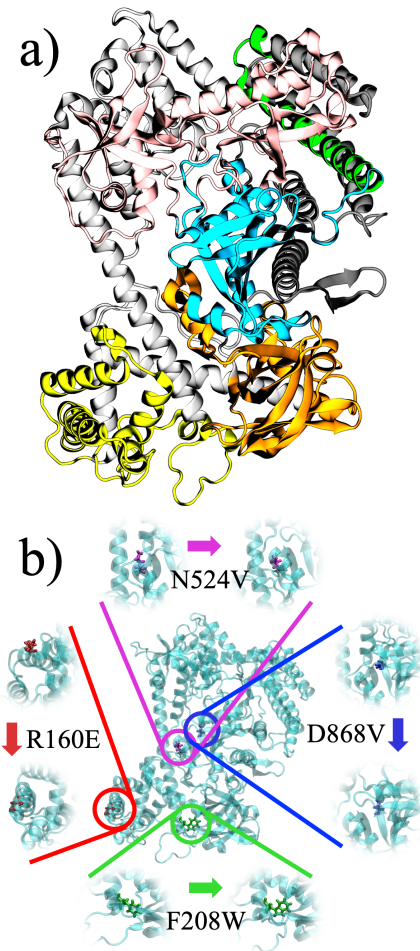

**Figure 1:** Schematic of (a) domain assignment of BrCas12b with residues colored according to domains; detailed legend can be found in the SI (Figure S1) and (b) location of mutated residues within the protein.

MD simulations were performed on two systems, wild-type (WT) and mutant (MT) BrCas12b, both in apo form (no nucleotide bound to the protein). The structure of WT BrCas12b was modeled using AlphaFold v2.3.1.[28] MT BrCas12b was also modeled using AlphaFold, with four point mutations (R160E/F208W/N524V/D868V) reported in a previous study by Nguyen et al.,[20] which outperformed other tested mutation combinations and displayed high-temperature stability and activity. Domain assignment (Figure 1a and S1) was made based on sequence alignment on the existing crystal structure of BthC2c1/BthCas12b (PDB: 5WTI) available in the RCSB PDB databank and domain organization proposed by Wu et al.[29]

All simulations were conducted using GROMACS 2021.3 software[30] with GPU acceleration. CHARMM36m[31] forcefield and TIP3P model was used to represent the protein and waters, respectively. Protein was solvated in a cubic box with the edge at least 2.0 nm from protein, resulting in box sizes ranging 15.6 to 17 nm with a total of ~ 380,000 to ~520,000 atoms based on the conformation of protein adopted. $Cl^-$ ions were introduced to neutralize each system. Hydrogen bonds were constrained during the simulation using the LINCS[32] algorithm. The cutoff for short-range electrostatic and Van der Waals interactions was set to 1.2 nm. The particle-mesh Ewald[33] method was used to compute long-range electrostatic interactions; fast Fourier transform grid spacing was 0.15 nm. Van der Waals interactions were smoothed between 1.0 to 1.2 nm using a force-base switching function.[34] Constant temperature was maintained using the Nosé – hoover[35] algorithm. For the constant-pressure simulations, the Parrinello-Rahman barostat (1 bar, 2.0-ps coupling constant) was used. The time step for numerical integration was 2 fs.

The systems, prepared as described above, were first subjected to steepest descent energy minimization with a maximum force of 1000 kJ·mol$^{-1}$·nm$^{-1}$, followed by two stages of 100 ps NVT equilibration. In the first stage, heavy atoms of the protein were restrained, and the solvent molecules were equilibrated. Initial velocities were randomly assigned according to Maxwell-Boltzmann distribution at 300 K. In the second stage, all atoms were equilibrated with the velocity adopted from the first stage and removed position restraint. Subsequently, a 1000 ps constant pressure and temperature (NPT) simulation was conducted at 300 K. The final outputs were used as the starting point for the production simulation.

With the protein structure obtained from AlphaFold, a 200 ns long simulation was first set up at 300 K to equilibrate the system further. The sampling of the simulation was confirmed with all-to-all root mean square deviation (RMSD) matrix (Figure S2) and was observed to have two distinct stages. To increase reproducibility and reliability, four replicates of each system were performed. The starting configuration of each replica was randomly selected from this trajectory between 100 ns – 200 ns. Following equilibration, a high-temperature unfolding simulation protocol was implemented, details can be found in our previous study.[22] Simulations were conducted at both 300 K and 400 K , for 200 ns. To further sample the configuration space and capture the essential dynamics of BrCas12b, one replicate of WT and MT BrCas12b simulations at both 300 K and 400 K was extended to 1 μs.

**Analysis**

Simulation trajectories were analyzed using GROMACS tools and visualized using VMD. The secondary structure of the protein and its evolution along the simulation were assigned using the inbuilt *timeline* tool of VMD with the STRIDE algorithm. The predicted circular dichroism (CD) spectra was generated by the PDBMD2CD web server using the resulting structure from each simulation trajectory.[36] Root mean square deviation (RMSD) of the whole protein was calculated by fitting to the backbone of the initial structure, whereas the per domain RMSD was calculated by fitting to the domains' initial backbone structure during the production run.

Principle component analysis (PCA) was done to capture the essential dynamics of the simulated systems. First, a covariance matrix of the Cα atoms was constructed by fitting the protein trajectory to the reference structure of protein backbone atoms using least-squares fitting. The resulting covariance matrix was diagonalized to obtain the eigenvectors as well as the corresponding eigenvalues. Each eigenvector represents a principal component (PC) that depicts a mode of movement, with the corresponding eigenvalue indicating the extent of the motion along the eigenvector. The process of arranging eigenvectors based on their eigenvalues reveals that the first principal component (PC1) aligns with the greatest amplitude motion of the system. The movement of the system along PC1 is commonly known as "essential dynamics".[37] In this work, we project the 1μs long trajectories in 2D defined by the first two principal components (PC1 and PC2) to characterize the conformational space sampled by simulation. The covariance matrix was constructed by using *gmx covar* and the projection was done using *gmx anaeig*, both commands inbuilt in GROMACS. Visualization of the protein motion was done using VMD.

The dynamic cross-correlation matrix (DCCM) was constructed to capture correlated and anti-correlated motions within BrCas12b. The covariance matrix constructed during PCA analysis was post-processed to generate the DCCM according to equation (1).[38]

$$CC_{ij} = \frac{\langle \Delta \vec{r_i}(t) \cdot \Delta \vec{r_j}(t) \rangle}{\left( \langle \Delta \vec{r_i}(t)^2 \rangle \langle \Delta \vec{r_j}(t)^2 \rangle \right)^{\frac{1}{2}}} \quad (1)$$

Where $\Delta r_i(t)$ represents the position of Cα atom at time t. The cross-correlation score ($Cs_i$) is calculated for each residue according to equation (2). The intra-domain score $Cs_i^{intra}$ is calculated when residues i and j belong to the same domain, and the inter-domain score $Cs_i^{inter}$ is calculated when they belong to different domains. Specifically, when residue i and j belong to different domains a and b, respectively, the calculated $Cs_i$ is the cross-correlation score coefficient of residue i with respect to domain b. In this way, the cumulative $Cs_i$ of all residues i belonging to domain a will give the cross-correlation between domains a and b.[23,39]

$$Cs_i = \sum_{i \neq j}^{N} CC_{ij} \quad (2)$$

**Results and Discussion**

**Global structure assessment**

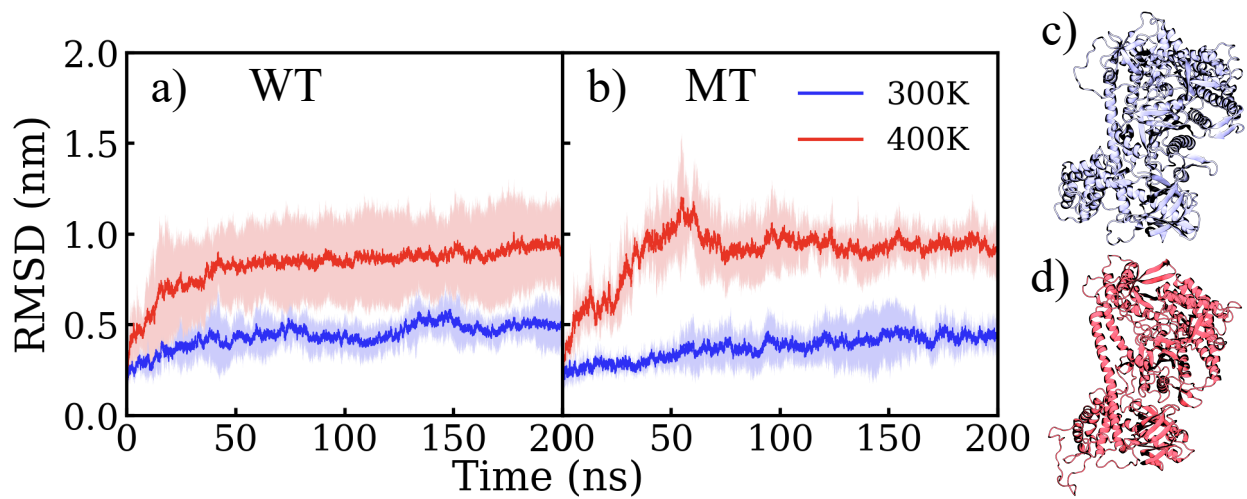

Figure 2: Backbone root mean square deviation (RMSD) of BrCas12b (a) wild type (WT) and (b) mutant (MT) as a function of simulation time, and snapshots of BrCas12b wild type at 200 ns from (c) 300 K and (d) 400 K. The solid lines indicate the mean RMSD value averaged over four independent trials, while the shaded region indicates the standard deviation.

To understand how point mutations enhance the thermostability of BrCas12b, we carry out MD simulations at 300 K and 400 K to capture the unfolding process of the WT and MT BrCas12b. Figure 2 shows the backbone root mean square deviation (RMSD). The RMSD was calculated by fitting the backbone carbon atoms and using the first frame of the simulation trajectory as the reference structure. At 300 K, we observe a relatively small increase of RMSD (< 5 Å) within 200 ns for both WT and MT, indicating a minor deviation from their initial configuration, considering the relatively large structure of the protein (Figure 2a and 2b). We also compute the change in hydrogen bonds and solvent accessible surface area (Supporting Information). Both WT and MT have a relatively stable hydrogen bond network, with an average total hydrogen bond number (HBN) of 840 (±15) at 300 K (Figure S3). At this temperature, the total solvent accessible surface area (SASA) of both WT and MT fluctuates at around 600 nm$^2$ (Figure S3). Secondary structure analysis (Figure S4) shows a small fluctuation of the helix and sheet components of both proteins, in agreement with the predicted circular dichroism spectra (Figure S5). The overall structural features of the apo BrCas12b for both WT and MT at room temperature are comparable, with a steady hydrogen bond network and corresponding SASA, and the structure deviates marginally from its starting conformation.

At 400 K, we see that the RMSD of both WT and MT increases with time compared to 300 K, as expected. This indicates a change in the protein structure at 400 K compared to its initial structure, with both proteins showing approximately the same RMSD of 1.0 nm over the last 100 ns of simulation time. Experimentally, it was observed that the MT protein's melting temperature is higher than that of the WT protein, and thus more thermostable.[20] However, the overall RMSD does not indicate a major difference in the thermostability of the two proteins. This could be because the experiments were performed at around 60 °C–65 °C (~335 K) while the simulations are conducted at 400 K for efficiency. Hydrogen bonds were observed to decrease for WT and MT but with a small change of 30 fewer hydrogen bonds at 400 K for both (Figure S3). Interestingly, SASA showed a decreasing trend at elevated temperatures for both WT and MT, with MT showing a slower decrease at the beginning, and a lower overall surface area at the end (Figure S3). The decrease in SASA indicates the collapse and shrinking of the protein structure at high temperatures, which differs from other proteins that we have investigated before, like insulin and lysozyme, that show an increase in SASA at elevated temperatures. Considering the thermophilic nature of the

BrCas12b, high temperatures trigger hydrophobic residues to come closer to prevent their exposure at the protein-solvent interface. The decreased SASA at high temperatures has been observed in other thermophilic proteins.[40–42] Secondary structure analysis (Figure S4) shows a decrease in helix and sheet components in both WT and MT, while predicted circular dichroism spectra (Figure S5) also indicate the loss of secondary structure, suggesting that parts of the protein are unfolding.

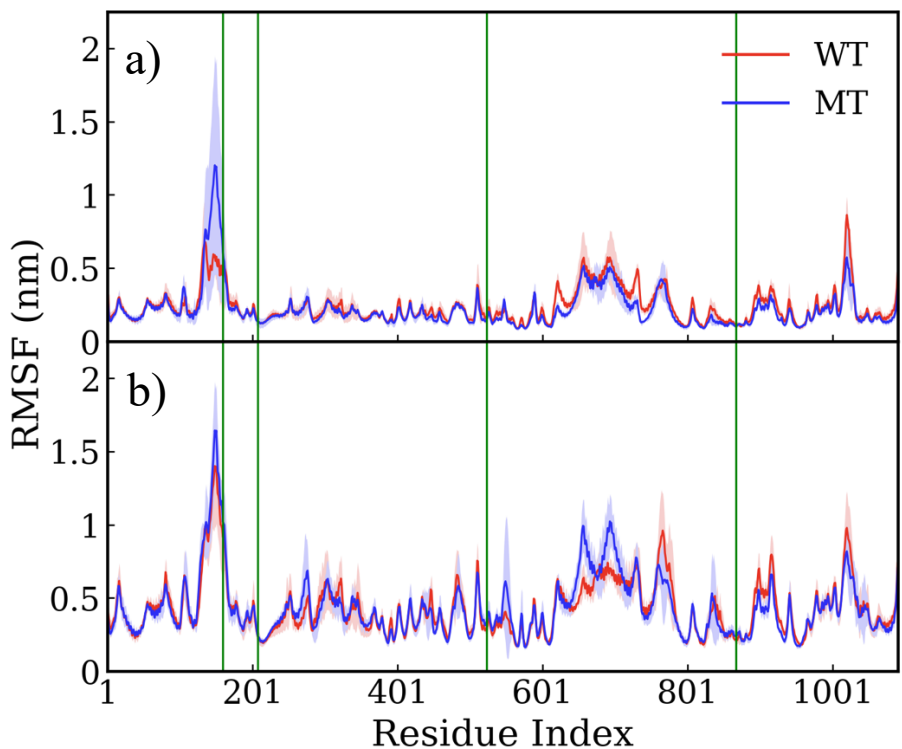

**Figure 3: Root mean square fluctuation (RMSF) of BrCas12b wild type (WT) and mutant (MT) from (a) 300 K and (b) 400 K, with four green vertical lines highlighting the mutation points R160E, F208W, N524V, and D868V. The solid lines indicate the mean value averaged over four independent trials, while the shaded region indicates the standard deviation.**

To understand the effect that the point mutations have on the flexibility of the residues, we analyzed the root mean square fluctuation (RMSF) on the backbone atoms of each residue; this is shown in Figure 3. At 300 K, the mutation of residue R160E leads to a significant increase in the fluctuations of the protein residues close to it, with an RMSF value of 1.2 nm for the mutated protein, compared to that of 0.6 nm for the wild-type protein. We ascribe this to the change of the residue charge from positive to negative after mutation. The remaining point mutations comprise hydrophobic side chains, leading to a slightly lower RMSF. At 400 K, the difference of RMSF between WT and MT was lower, with residue R160E having a marginally larger value in MT compared to WT. Given the relatively large protein size, it is hard to discern structural changes brought about by just four mutations (effectively 0.4% of the entire protein) using these metrics alone. Detailed analyses at a local residue level can shed more light on the underlying mechanism of stabilization stemming from the mutations, which we discuss below.

**Structural analysis of mutated domains**

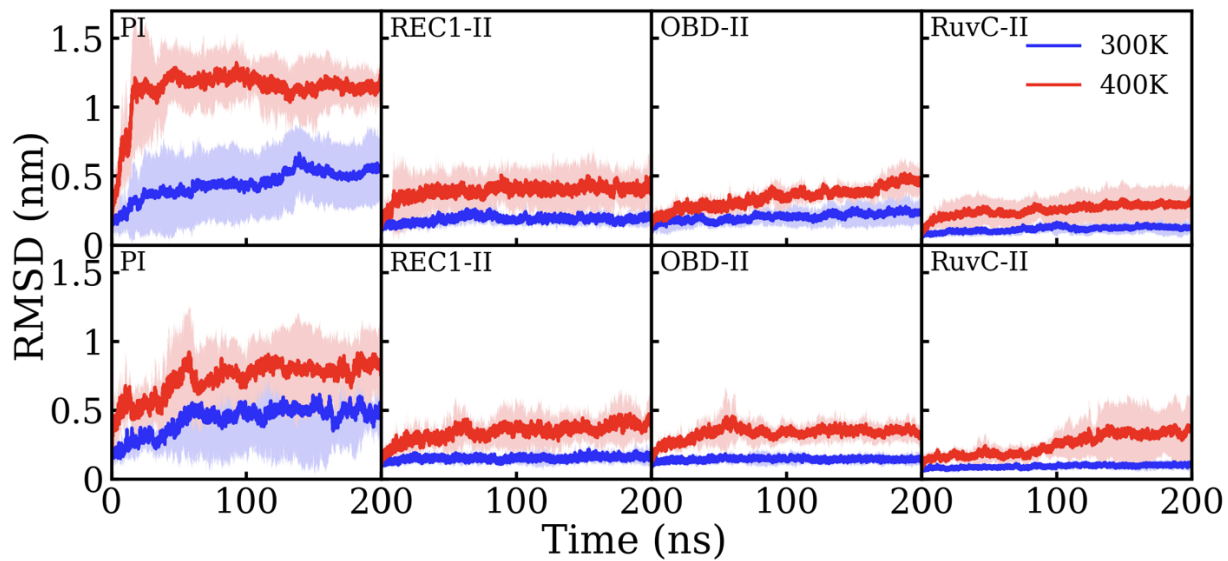

**Figure 4: RMSD of four domains which contain point mutations in the wildtype (top) and mutant (bottom) protein, as labeled. The solid lines indicate mean value averaged over four independent trials, while the shaded region indicates the standard deviation.**

To understand the effect of high temperature on the stability of MT and WT BrCas12b, we decompose the protein into domains as discussed in the Methods section. Particularly, the BthC2c1 structure features a two-lobed composition, with an α-helical recognition (REC) lobe and a nuclease (NUC) lobe. The REC lobe is formed by a PAM-interacting (PI) domain, REC1 and REC2 domains, and an extended α helix known as the bridge helix (BH). In the NUC lobe, there is an OBD domain, a RuvC domain, and a domain with undetermined functions, referred to as the "UK" domain. The RuvC domain in the NUC lobe, which is composed of three separate RuvC motifs (RuvC I-III), interacts with the REC2 domain in the REC lobe. The connection between the RuvC domain and the REC1 domain is primarily facilitated through the UK domain.[29] The domains with mutated residues are PI, REC1-II, OBD-II, and RuvC-II.

RMSD analysis was carried out for each decomposed domain. Domains with mutated residues are highlighted in Figure 4, and the RMSD for all domains are presented in Figure S6.

RMSD is calculated by first performing root-mean-square fit on backbone atoms of a specific domain to its initial structure. At 300 K, the RMSD of most domains in both WT and MT are low, indicating small deviations from their initial structure. Exceptions to this are domains PI (Figure 4), REC2, and UK-II (Figure S6), in which the RMSD increases over time compared to the other domains. The higher RMSD for these three domains even at 300 K indicates that they are more likely to undergo conformational changes compared to the other domains, especially PI, which shows an RMSD near 5 Å in the last 50 ns of simulation time. The higher RMSD of these domains also coincides with the higher RMSF, indicating higher structural flexibility. Previous studies on other CRIPSR-Cas proteins like Cas9 and Cas12a suggest the high flexibility of PI domain and the C-term (part of UK-II here), is substantially reduced after being bound to nucleic acid.[23,24] At 400 K, the trend of RMSD is different across multiple domains. At 400 K in WT, there are several domains, including REC1-I, BH, and RuvC-III (Figure S6), which do not undergo dramatic conformational change and show a comparable RMSD to 300 K. The RMSD of domains OBD-I (WED-1), REC1-II, OBD-II, RuvC-I, RuvC-II, and UK-I show a slight increase but do not exceed 5 Å. Domains PI, REC2, and UK-II, which show high conformational change at even 300 K have the highest change at 400 K as well. Of particular importance is domain PI, which shows a dramatic structural change at the beginning of the high-temperature simulation, indicated by the steep slope of the RMSD, before 20 ns. After the mutation, no apparent difference was observed in most domains except for domain PI (Figure 4). Though there is a sharp rise in RMSD at the start of the simulation, the RMSD plateaus at ~0.9 nm for MT, lower than the value for WT which is ~1.3 nm, suggesting a smaller change in the MT structure compared to the WT structure. As for the other domains which contain mutated residues, REC1-II and RuvC-II show a slower increase of RMSD than their WT counterpart, whereas domain OBD-II shows a lower RMSD for the last 40 ns of simulation.

Table 1: Hydrogen bond occupancy of hydrogen bonds formed between mutated residues to other residues. Mutated residues are highlighted in bold. Hydrogen bond pairs with >10% occupancy are shown in the form of donor(atom)-acceptor(atom), with occupancy averaged over four 200 ns trajectories.

| Donor(atom) – acceptor(atom) (WT) | 300 K | 400 K | Donor(atom) – acceptor(atom) (MT) | 300 K | 400 K |
|---|---|---|---|---|---|
| **R160**(HN)-R156(O) | 89.5 | 41.2 | **E160**(HN)-R156(O) | 83.5 | 49.6 |
| **R160**(H11)-E157(OE2) | 11.8 | 5.4 | A164(HN)-**E160**(O) | 70.1 | 10.6 |
| **R160**(H11)-E157(OE1) | 11.9 | 5.8 | | | |
| M212(HN)-**F208**(O) | 99.2 | 93.0 | M212(HN)-**W208**(O) | 99.5 | 97.5 |
| H394(HE2)-**F208**(N) | 11.1 | 2.2 | **W208**(HE1)-E422(OE1) | 94.2 | 48.0 |
| **N524**(HN)-K530(O) | 98.7 | 95.8 | **V524**(HN)-K530(O) | 98.9 | 93.3 |
| K530(HN)-**N524**(O) | 78.9 | 77.0 | K530(HN)-**V524**(O) | 94.8 | 60.8 |
| K526(HN)-**N524**(OD1) | 66.5 | 31.2 | | | |
| K530(HZ1)-**D868**(OD1) | 54.5 | 54.7 | **V868**(HN)-V531(O) | 84.8 | 83.1 |
| **D868**(HN)-V531(O) | 56.9 | 45.6 | | | |
| K530(HZ1)-**D868**(OD2) | 67.0 | 51.2 | | | |

To further understand intra-protein structural changes, we investigated the formation of hydrogen bonds, to capture how mutation alters the original hydrogen bond network of the protein. Table 1 and Figure S7 show the formation of hydrogen bonds between the mutated reside and other protein residues before and after mutation. As the mutations contribute both to the formation or disappearance of hydrogen bonds, a comparable number of total hydrogen bonds before and after mutation at 300 K is observed. We then investigate residues pairs that give high occupation values (i.e., hydrogen bonds observed in more than 10% of simulation time). For R160, the hydrogen bonds are formed mainly with R156 (backbone oxygen) and E157 (side chain oxygen). Specifically, R160 – R156 was found to be the dominant hydrogen bond with a high occupation (89%) on average across all four replicates. This is an example of a hydrogen bond between pairs that are less than four residues apart and is integral in maintaining the alpha-helical structure of the PI domain. After mutation (R160 → E160), A164 takes the place of E157 as an acceptor to form a hydrogen bond with E160 and a larger helix bundle is observed. For F208, residue pairs forming hydrogen bonds are M212 - F208 and H394 - F208, where F208 is the acceptor. M212 - F208 has a relatively high occupation, 99% on average. After the mutation (F208W → W208), the hydrogen bond previously formed between H394 and F208 is no longer observed between H394 and W208. W208 becomes the donor to form a hydrogen bond with E422, which possesses a steadier interaction with an occupation of ~94%. A closer examination of the switch of hydrogen bond involving F208W suggests the change of interaction with H398, located in a loop, to E422, located in a bundle of beta sheets. The formation of W208 - E422 thus strengthens the interaction between the helix bundle of REC1-II and the beta-sheet of OBD-II. Unlike R160E and F208W, N524V and D868V lead to a decrease in the number of hydrogen bonds formed to other residues due to the change of the long side chains to relatively short and hydrophobic methyl side chains.

Post mutation of N524 to V524, its hydrogen bond with K526 disappears, which was located on the loop connecting two anti-parallel beta-sheets. D868 also forms hydrogen bonds with K530 and V531. Mutation of D868 → V868 results in the loss of hydrogen bonds with K530 but hydrogen bonds between hydrophobic residues such as V531 is maintained. We posit that the diminished structural change at high temperature for specific domains alongside the change in the hydrogen bond network results in the increased stability and activity at higher temperature.

**Protein essential dynamics**

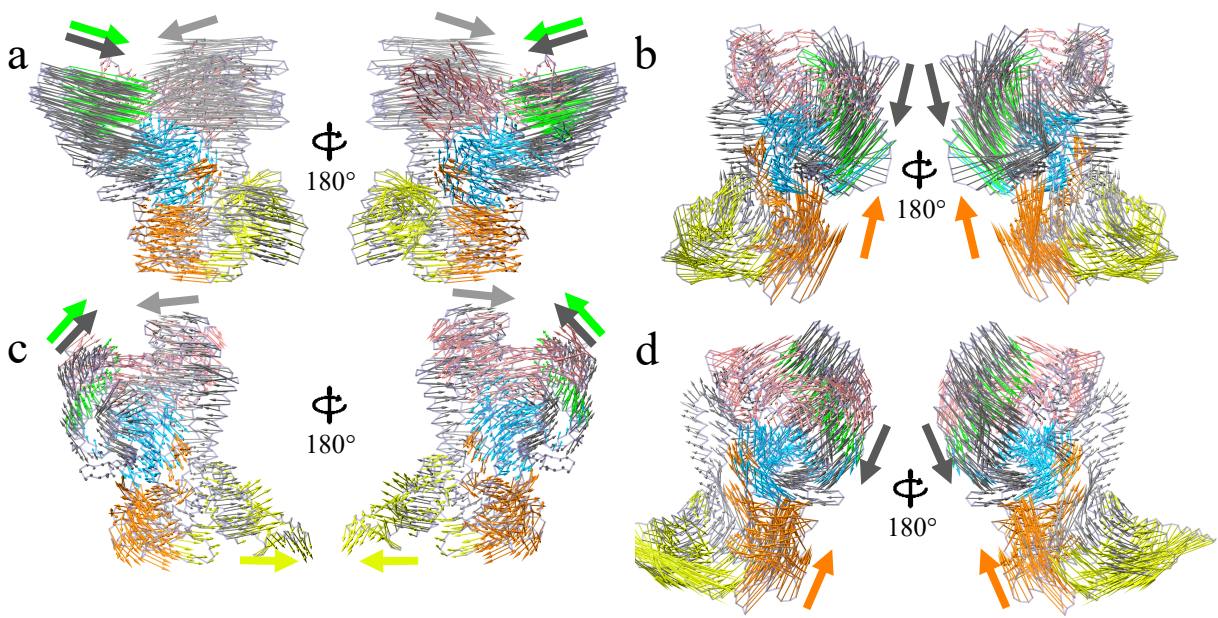

**Figure 5: Essential dynamics derived from the first principal component (PC1) of wild-type BrCas12b at (a) 300 K, (b) 400 K, and mutant BrCas12b at (c) 300 K, (d) 400 K shown using arrows of sizes equivalent to the amplitude of motions, with colors adapted from Figure 1 to distinguish motions of different domains.**

The mutations bring structural change to BrCas12b, potentially responsible for the improved stability and activity. To decouple the complex motion of the apo-BrCas12b protein and to understand the essential degree of freedom for both WT and MT, we performed PCA, as described in the Methods section. The dynamics of the protein along the first principal component is usually referred to as essential dynamics.[37]

Upon projecting the simulation trajectory on the top two principal components (PC1 and PC2), we examine the projected motion of each principal component (Figure 5 and S8). For WT BrCas12b at 300 K, the essential dynamics presented by PC1 (Figure 5a) is the large-scale movement between domains BH, REC2, and domain REC1-II toward each other, alongside relative motions between domain PI and domain OBD-II with a smaller amplitude. The motions along PC1 were previously characterized to be "open" and "close" of the CRISPR-Cas protein to accommodate the nucleic acid, including Cas9 and, more recently, Cas12a.[23,24] Herein, BrCas12b also possesses the "open" and "close" motion, potentially accommodating the formation of complexes with RNA or DNA. The projected motion on PC2 highlights the orthogonal motion of domain BH together with REC2 compared to PC1, presenting a twisting motion making the front part of domain RuvC-I act as a fulcrum. A significant amplitude motion is also observed for domain PI, with an alpha-helix bundle moving inward and outward around the hollow part of the apo-BrCas12b. With the four RFND mutations, a smaller 'open' and 'close' motion is observed for MT BrCas12b from the projection of PC1 (Figure 5c). The previously observed significant amplitude movement between domain BH, REC2, and domain REC1-II was much smaller than WT BrCas12b. The initially small-scale motion from domain PI and domain OBD-II is enhanced in MT BrCas12b, with PI presenting a relative open conformation, where the point mutation was introduced. For PC2, MT presents a similar motion for domain BH and REC2, as seen in WT. The motion of PI, however, toward the outside of the protein is different due to the open conformation adopted.

We also performed PCA on high-temperature simulations of the WT and MT BrCas12b, to understand how the mutation alters the essential dynamics at high temperatures and whether the mutation changes the mode of motion that unfolds the protein. An inward motion is observed for most of the domains from PC1, primarily domains REC2 and BH, moving together with domains OBD-I and II toward each other. This motion results in a lower exposed surface area, which also explains the downward trend of the SASA as the simulation proceeds. The gap in the middle of the protein, potentially to accommodate nucleic acids, also collapses. As for PC2, the domain BH and domain REC2 complex move towards domain REC1-II, however, domain REC1-II moves away from BH and REC2. The extent of this motion is small, which preserves the open and close motion observed at 300 K. The same motion was observed with the mutation between domain REC2, BH, and domain OBD-I, II. Domain PI also preserves the outward trend at 300 K and undergoes a large-scale motion toward the OBD-II than WT. The "open" motion is also observed for MT at 400 K, together with the inward motion of domain OBD, which differs from the WT protein. The domain PI then undergoes a significant outward expansion, which is not observed for in WT. By looking at the conformational space presented by the projection onto PC1 and PC2, we can clearly identify the open and closed motion along the PC1 (Figure S9). Covariance analysis showed that trace values of WT and MT are 483 nm$^2$ and 162 nm$^2$, respectively, at 300 K. The lower trace value of MT indicates a generally lower flexibility post mutation. At 400 K, the trace values of WT and MT are 376 nm$^2$ and 280 nm$^2$, respectively.

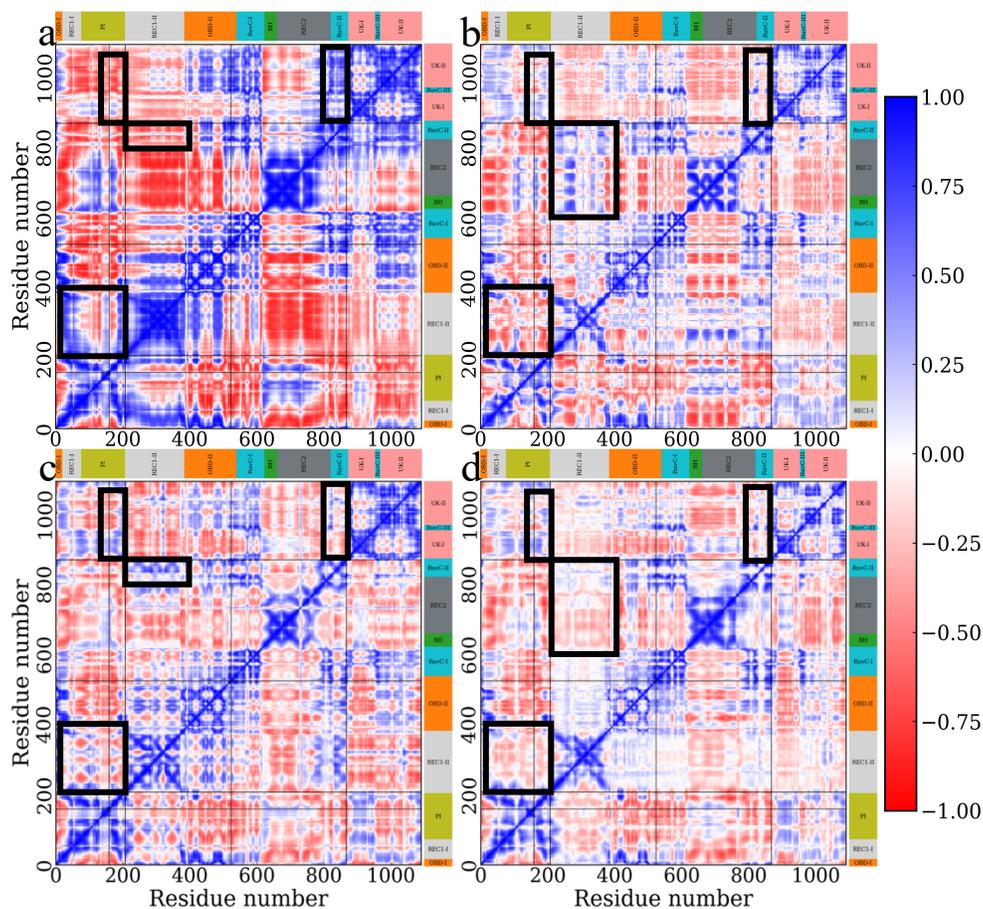

**Figure 6 Dynamic cross-correlation matrix (DCCM) calculated from 1μs simulation trajectory of wild-type BrCas12b at (a) 300 K, (b) 400 K, and mutant BrCas12b at (c) 300 K, (d) 400 K, with correlation between pairs from negative to positive colored respectively red to blue. The protein sequence is shown along the x- and y-axis with the corresponding domain assignment highlighted with colors. The mutated residues are highlighted in the black boxes.**

**Cross correlation changes upon mutation**

Next, we performed a cross-correlation analysis to probe the correlation of individual residues and corresponding protein domains. The results are shown in Figure 6, with the positive value of C(i,j) implying positively correlated movement whereby the two Cα atoms moved in the same direction, indicating a coordinated or coupled motion. In contrast, the negative value indicates anti-correlated movements whereby the atoms move in the opposite direction.[38] As shown in Figure 6a and 6c, at 300 K, the introduction of point mutations changes the correlated motions nearby. For R160E and F208W, the mutation slightly changed the correlated motion from positive to negative, involving part of domain PI and REC 1-II. The same correlation change is observed between domain RuvC-II, where D868V is located, and domain UK-I, UK-II. The shift from negative to positive was then observed from the correlation between domain RuvC-II and domain REC1-II as well as domain PI and domain UK-I, UK-II. At 400 K, the initial high correlation was observed to decrease, especially for the off-diagonal component (Figure 6b and 6d). The correlation between domain PI and domain UK shows a change from negative to positive, similar to that of WT 300 K. The same result was also observed for the correlation between REC1-II and RuvC-II, domain RuvC-II, and domain UK. The correlation change between domain REC1-II and PI was rather complicated, with the reverse of correlation mode happening differently. However, the correlation changes in MT at 400 K mainly decreased instead of completely changing. Thus, the introduced point mutations change the correlation mode at 300 K and reduces the amplitude of correlation change at 400 K.

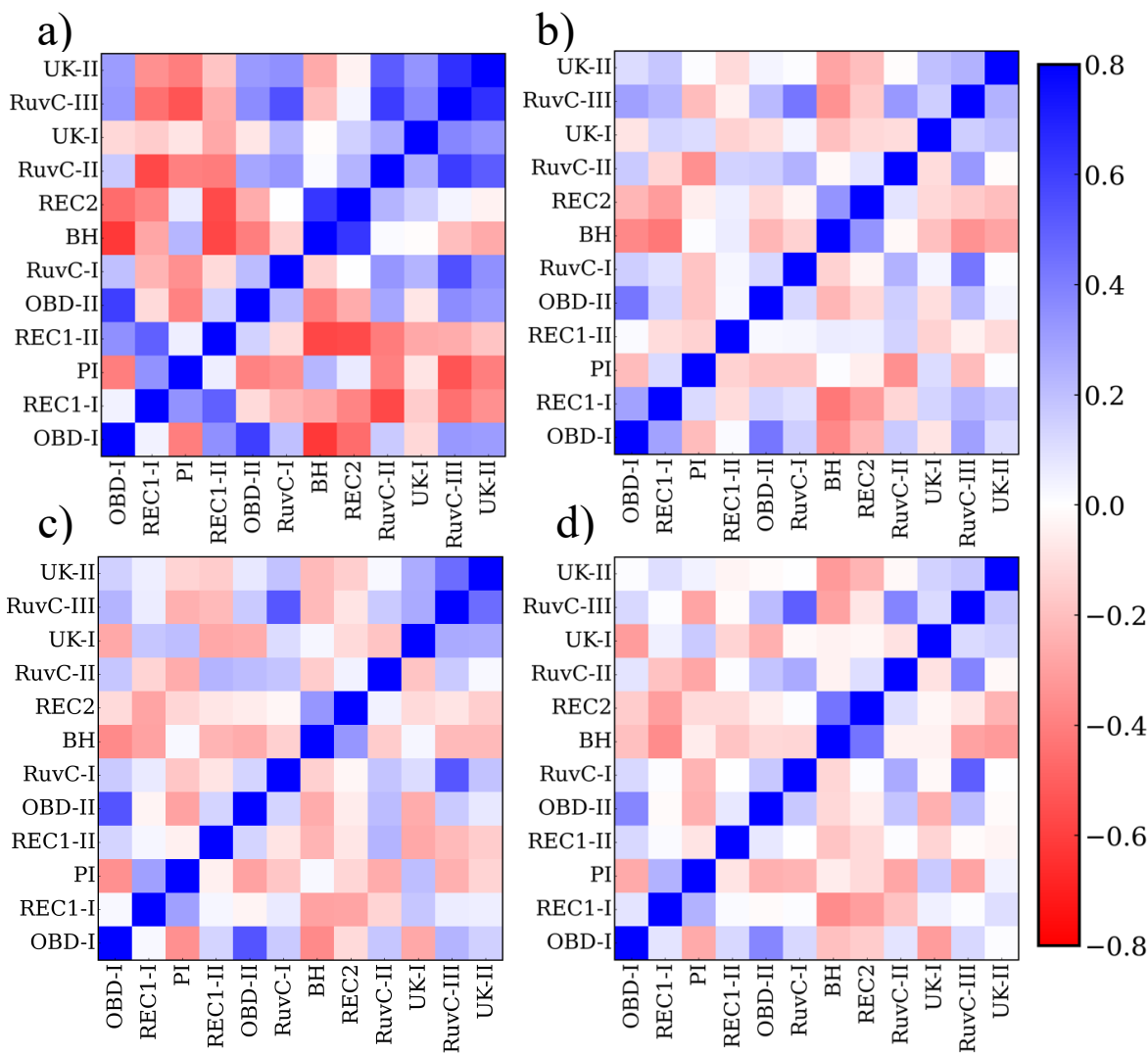

**Figure 7** Accumulated residue correlation between separate domains for wild-type BrCas12b at (a) 300 K, (b) 400 K, and mutated-type BrCas12b at (c) 300 K, (d) 400 K.

Finally, per residue correlation was calculated to highlight the interdomain correlation. As shown in Figure 7a, strong positive and negative correlations were observed across multiple domains for WT at 300 K. As mentioned earlier, large-scale 'open' and 'close' motions between domain REC1-II and domain BH, REC2, can also be identified here to show a strong negative correlation, indicating the opposite direction movement, together with the smaller amplitude motion between domain PI and domain OBD-II. Those motions become smaller after introducing the four mutations, as shown in Figure 7c. At high temperatures, initially strongly correlated motions between specific domains are observed to change (Figure 7b and 7d), especially for domain REC1-I and REC1-II. The initially strong negative correlations are observed to decrease or even reverse to positive correlations. However, the correlation between REC1-I and REC1-II changed from positive to slightly negative, indicating the change from moving toward the same direction to the opposite direction. For MT, the high temperature does not change the pattern of cross-correlation much. However, it is worth noting that the dynamic cross correlation may only account for the linear correlation.

**Conclusions**

We conducted MD simulations on the wild type (WT) and mutated (MT) BrCas12b at low and high temperatures, revealing the unfolding behavior of both WT and MT as well as the change of essential dynamics upon mutation to understand the origins of thermostability in mutant BrCas12b. The global property assessment on BrCas12b variants at ambient and elevated temperatures helped identify overall structural change, which indicates the unfolding of the protein at high temperatures. Further analysis on the decomposed domains based on sequence alignment of the protein highlighted the stabilization of the PI domain post mutation, while the other domains containing mutations show smaller structural change at the end of the simulation compared to the wild type. Hydrogen bond analysis shed light on the influence of mutations in generating new hydrogen bond networks, contributing to improved protein stability. The analysis of the essential dynamics from principal component analysis illustrates how wild type BrCas12b behaves at room temperature with the "open" and "close" motions observed in previous studies of other CRISPR-Cas proteins and how high temperature alters these dynamics to unfold the protein by first inducing the collapse of the initial structure followed by shrinking of the protein. Similar analysis of mutant BrCas12b suggests alternations in the protein dynamics after mutation, especially domain PI, as well as a change in the dominant motion during the unfolding at high temperatures. The cross-correlation between atom pairs and accumulated cross-correlation score between domains further corroborates the "open" and "close" motion and show how mutation decreases the correlation at both 300 K and 400 K. Overall, MD simulations of WT and MT BrCas12b at low and high temperature provide important details on how mutations improve the protein stability while altering the essential motions, which explains prior experimental observations. While the current study was done on apo-BrCas12b, future work on the protein complex formed with DNA/RNA will be conducted to understand the influence of point mutations on nucleotide binding in BrCas12b.

**Acknowledgement**

J.S acknowledges startup funds provided by the Department of Chemical Engineering and the Herbert Wertheim College of Engineering at the University of Florida (https://www.che.ufl.edu). J.S and Y.J gratefully acknowledge funding provided by the Oak Ridge Associated Universities (ORAU) Ralph E. Powe Junior Faculty Enhancement Award. P.K.J and S.R.R acknowledge support through the National Institute of General Medical Sciences (R35GM147788). K.H. acknowledges support by the National Science Foundation under Grant No. 1852111. Y.J and K.H acknowledge University of Florida Research Computing for providing computational resources and support (https://www.rc.ufl.edu/).

**Support Information**

Domain assignment details. Additional protein structure analysis includes total hydrogen bond number, total SASA, predicted CD spectra, backbone RMSD for different protein domains, and number of hydrogen bonds formed before and after mutation. Additional protein dynamics analysis includes protein dynamics projection onto PC2, simulation trajectory projected onto PC1-PC2 phase space, and cumulative contribution of all principal components.

Supporting Information for

**Exploring the Thermostability of CRISPR–Cas12b using Molecular Dynamics Simulations**

Yinhao Jia[1], Katelynn Horvath[2], Santosh R. Rananaware[1], Piyush K. Jain[1,3,4], Janani Sampath[1,*]

[1]Department of Chemical Engineering, University of Florida, Gainesville, FL, USA

[2] Department of Chemical and Biomolecular Engineering, University of Connecticut, Storrs, CT, USA

[3] Department of Molecular Genetics and Microbiology, College of Medicine, University of Florida, Gainesville, FL, USA

[4] Health Cancer Center, University of Florida, Gainesville, FL, USA

* jsampath@ufl.edu

**Table of content**

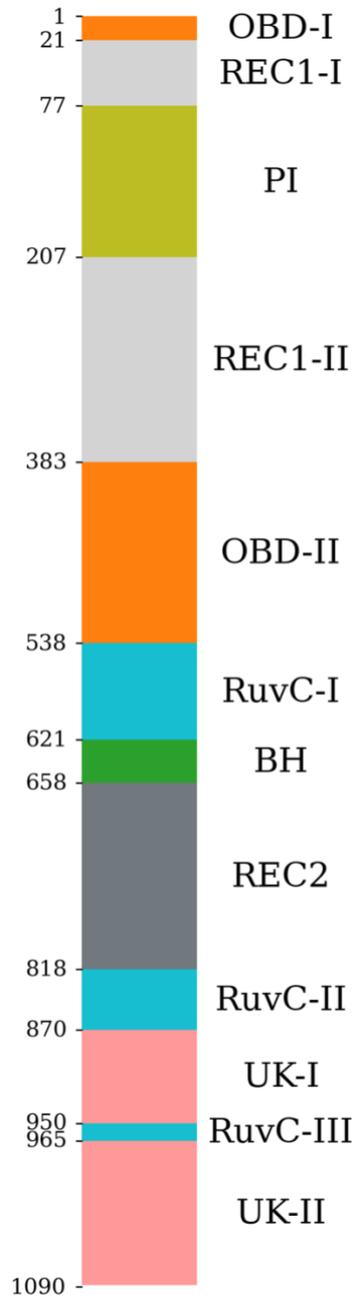

**Figure S1. BrCas12b domain assignment.**

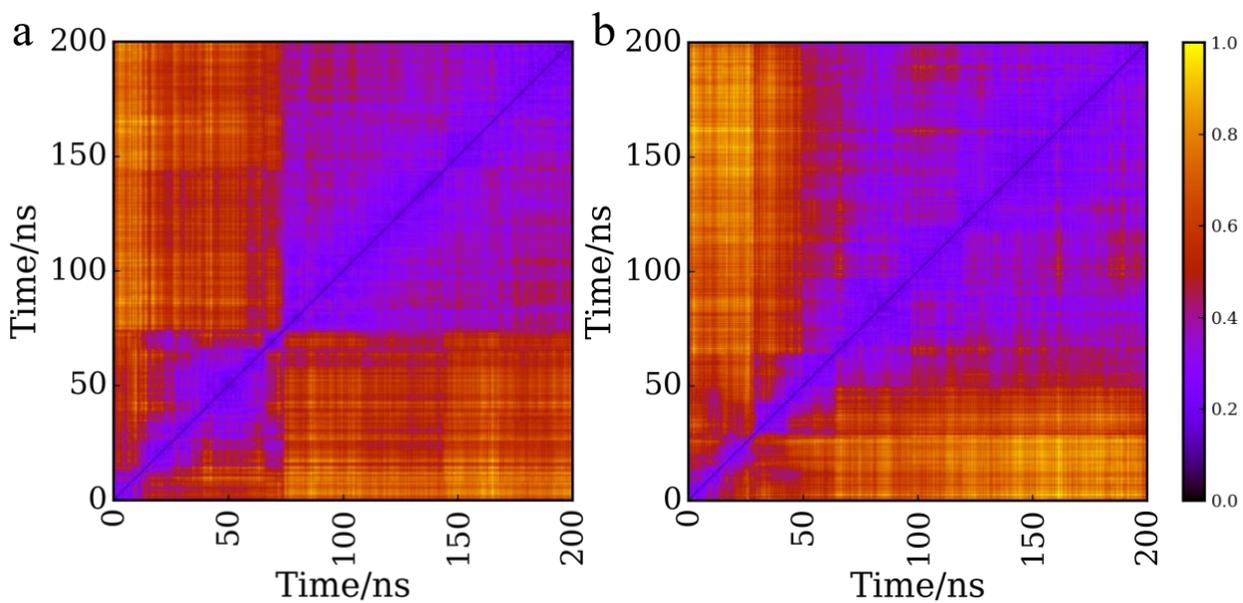

**Figure S2.** All to all RMSD of (a) wild-type BrCas12b and (b) mutated-type BrCas12b for the 200ns equilibration before random picking of configuration for the final simulation.

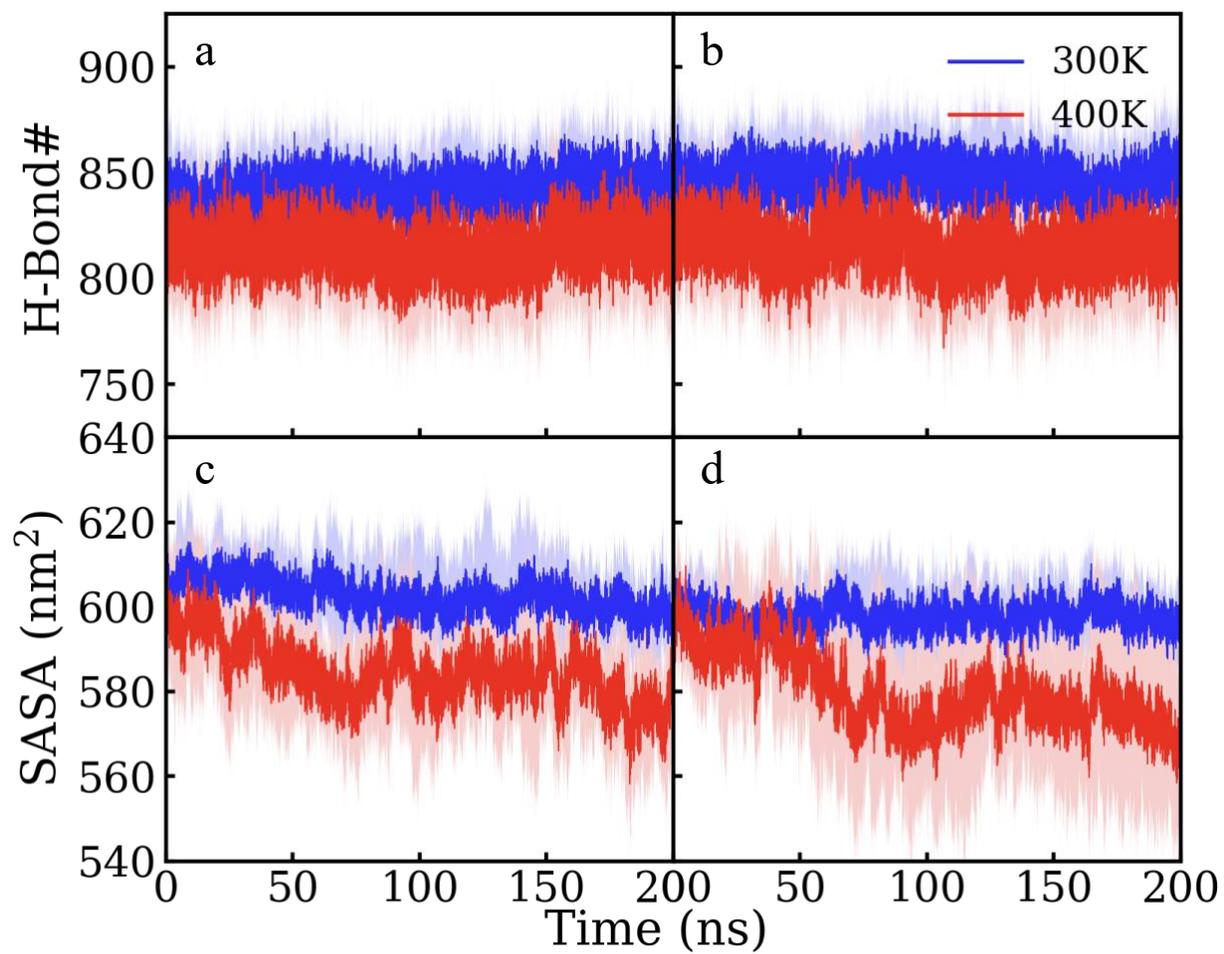

Figure S3. Intra protein hydrogen bond number (a, b) and total solvent accessible surface area (SASA) (c, d) of BrCas12b (a, c) wild type and (b, d) mutant type as a function of simulation time. The solid lines indicate values averaged over three independent trails, while the shaded region indicates the error calculated from the four replicas.

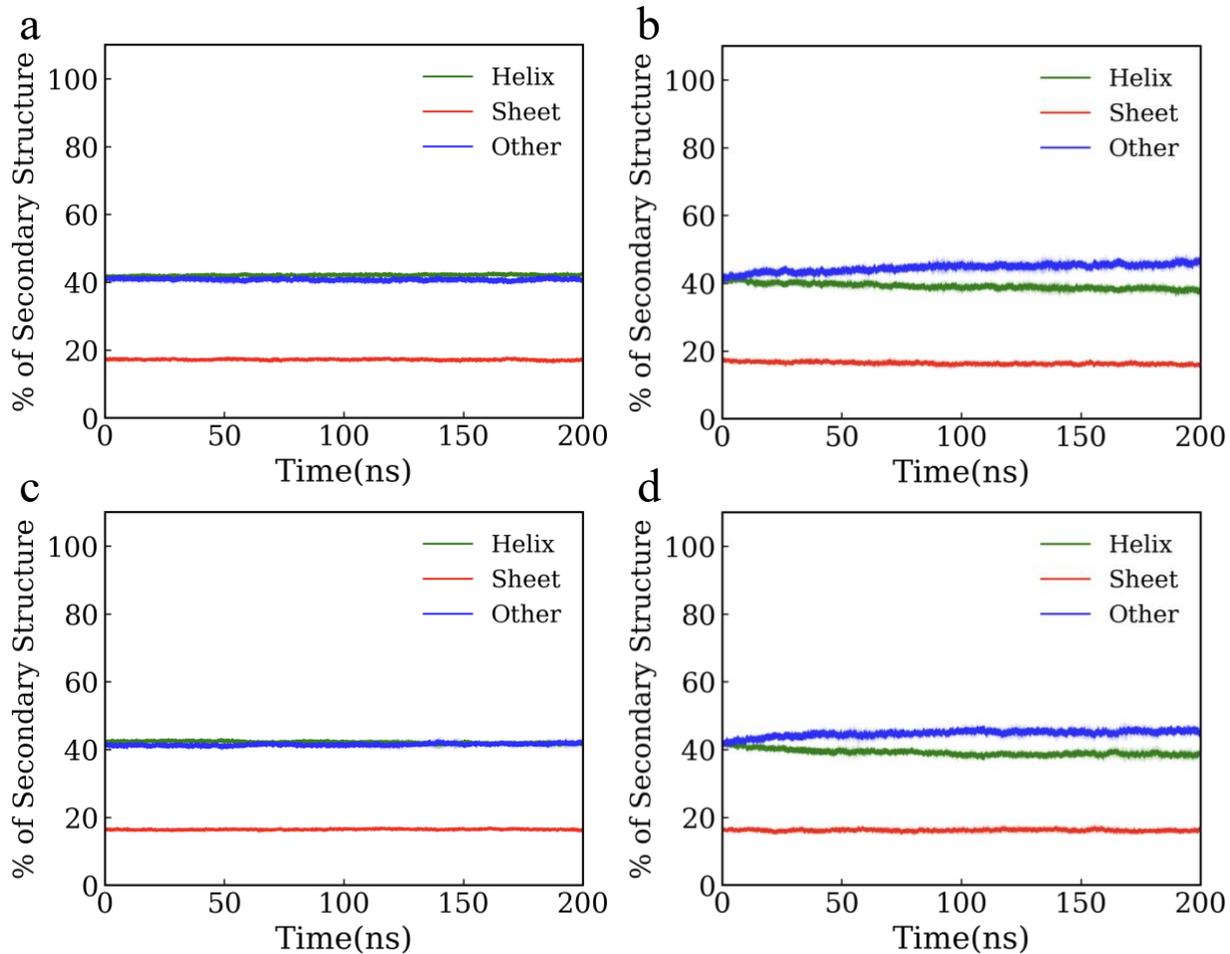

**Figure S4.** Secondary structure fraction of wild-type BrCas12b at (a) 300 K, (b) 400 K, and mutated-type BrCas12b at (c) 300 K, (d) 400 K. The solid lines indicate values averaged over three independent trails, while the shaded region indicates the error bar calculated from the three replicas.

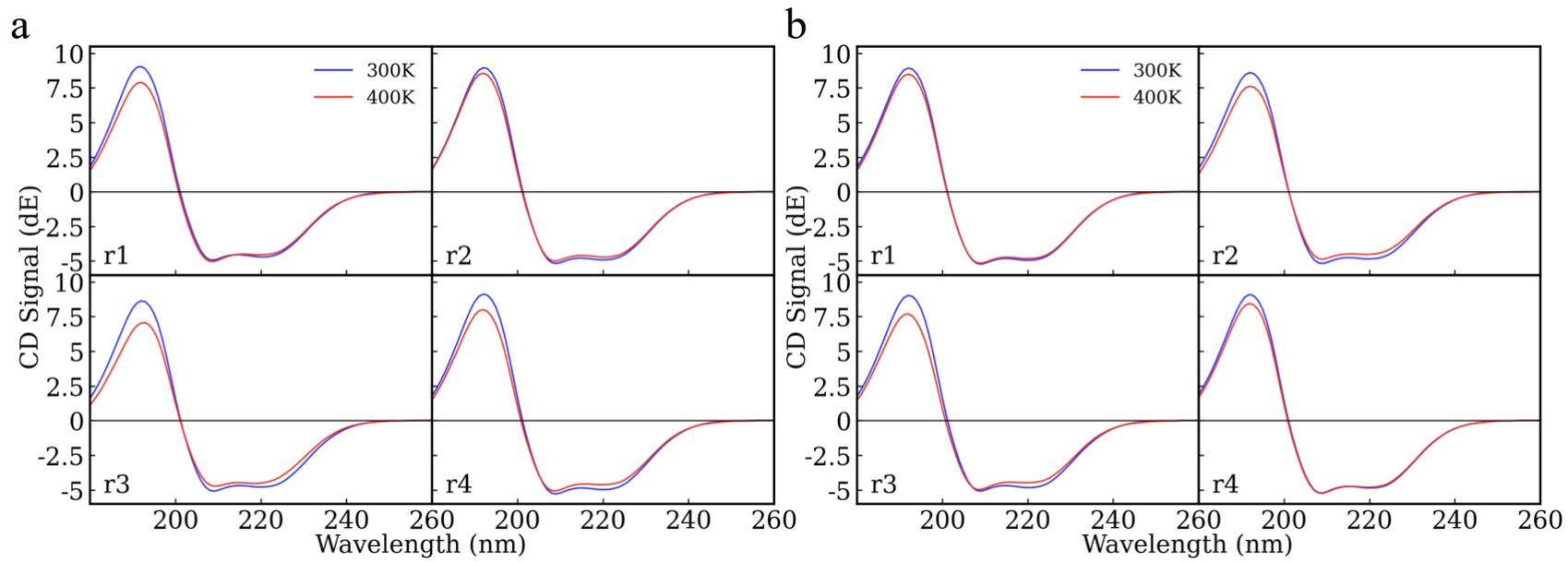

**Figure S5.** Predicted circular dichroism (CD) spectra of (a) wild-type BrCas12b and (b) mutated-type BrCas12b at 300 and 400 K after 200 ns simulation for four replicates (r1 – r4).

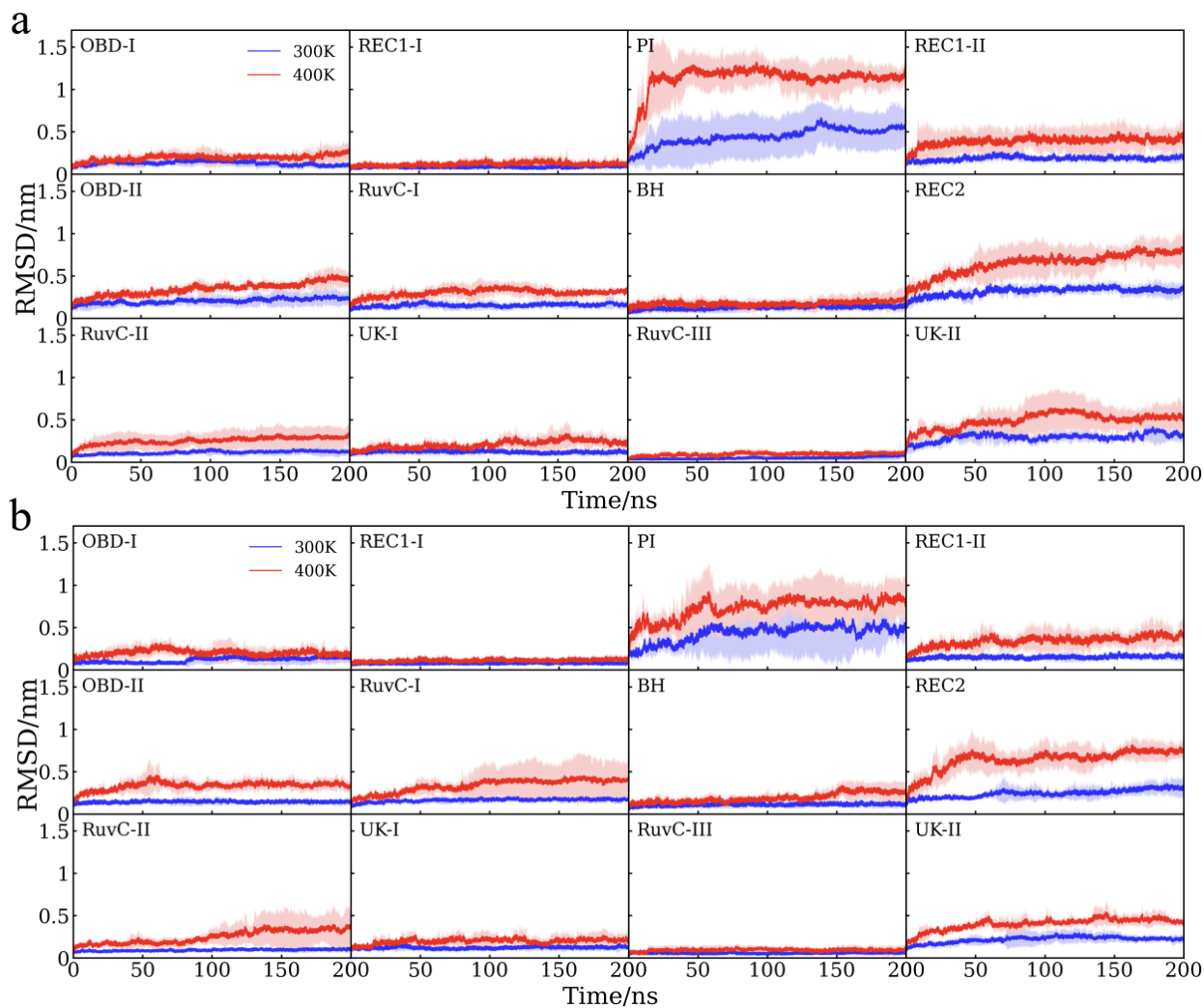

Figure S6. Backbone root mean squared deviation (RMSD) of each domain from (a) wild-type BrCas12b and (b) mutated-type BrCas12b, with the names of the domains shown at the top left corner. The solid lines indicate values averaged over three independent simulations, while the shaded region indicates the error bar calculated from the three replicas.

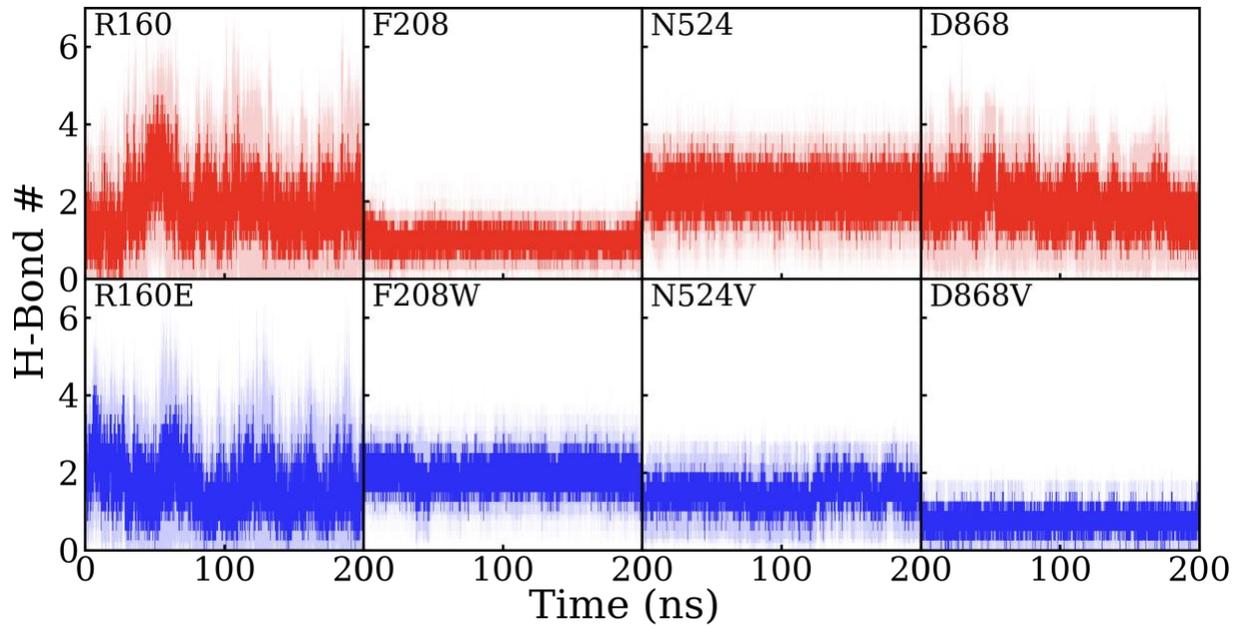

**Figure S7.** Number of hydrogen bond formed between the mutation point to other residues at 400 K, with each mutation point shown on the top left corner, as a function of simulation time. First row from wild-type, second row from mutated-type BrCas12b. The solid lines indicate values averaged over three independent trails, while the shaded region indicates the error bar calculated from the four replicas.

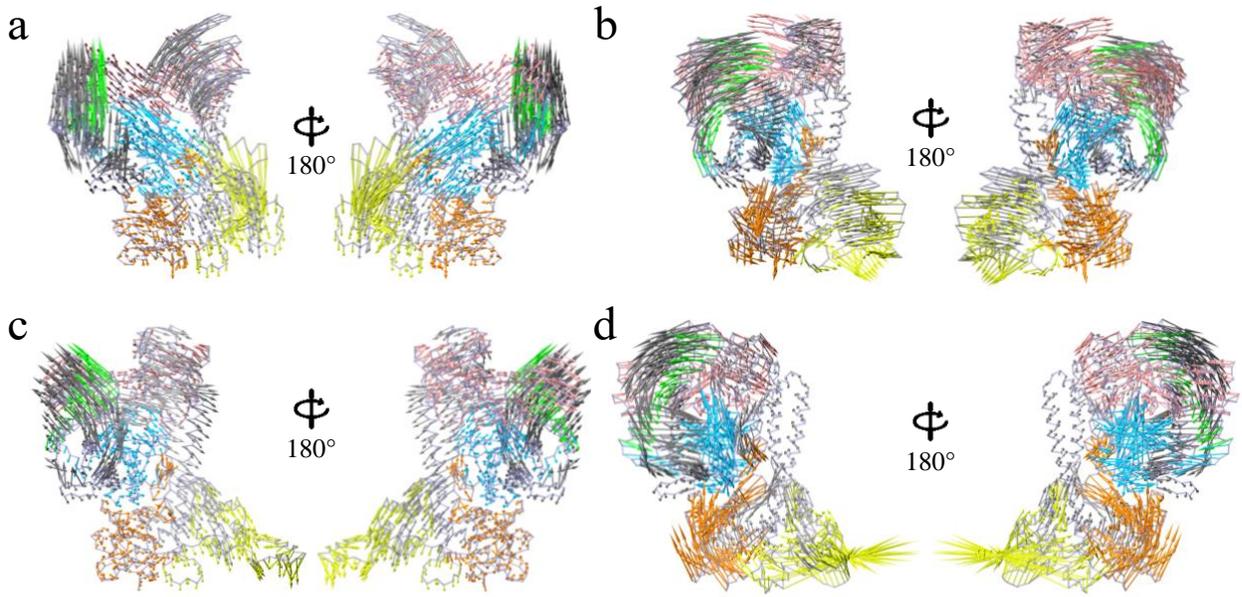

**Figure S8.** Motions obtained from principal component 2 (PC2) of wild-type BrCas12b at (a) 300 K, (b) 400 K, and mutated-type BrCas12b at (c) 300 K, (d) 400 K shown using arrows of sizes equivalent to the amplitude of motions, with colors adapted from Figure 1 to distinguish motions from different domains.

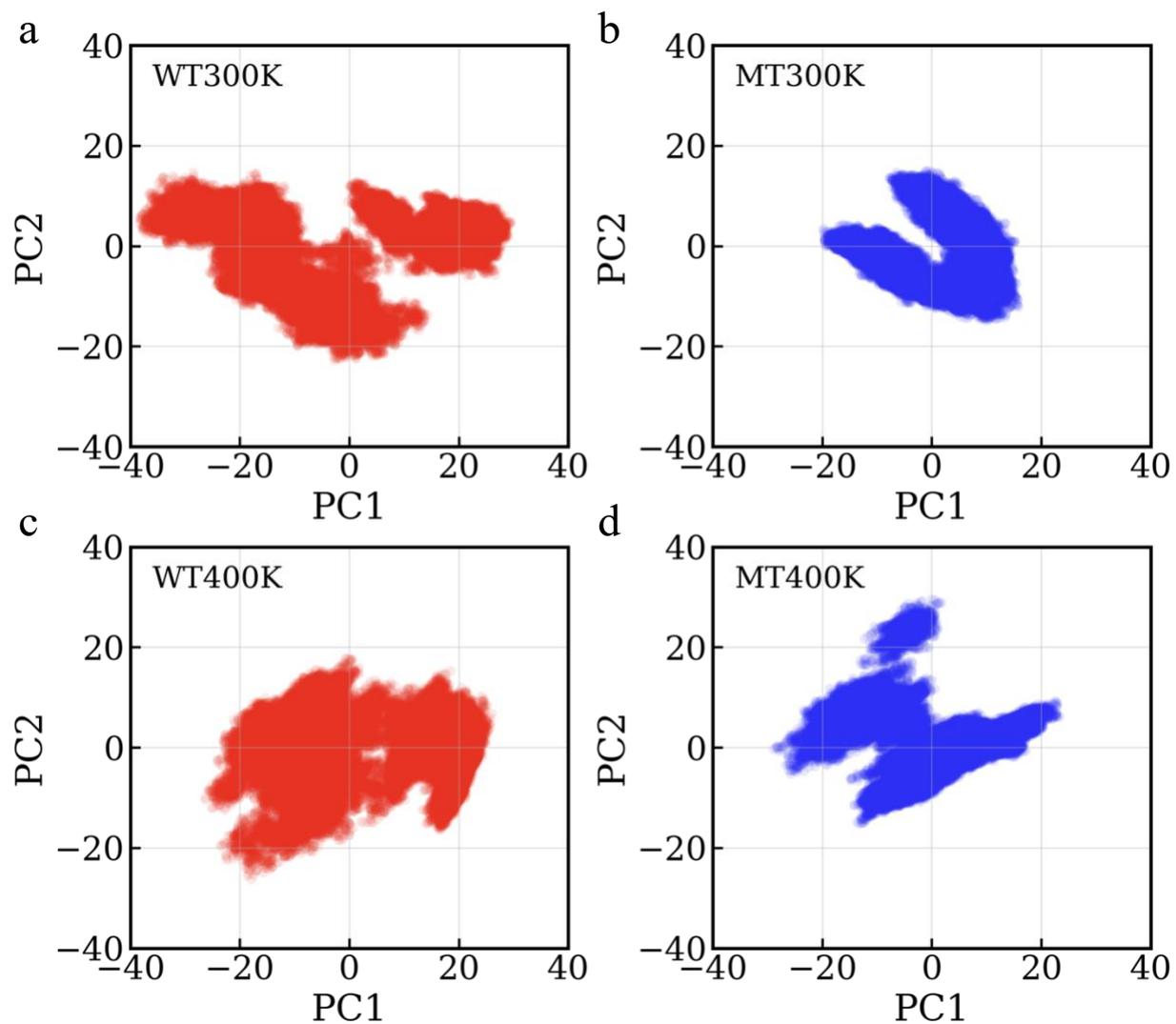

**Figure S9.** Projections of the simulation trajectory onto first and second principal component derived from (a) wild type BrCas12b at 300 K, (b) mutated type BrCas12b at 300 K, (c) wild type BrCas12b at 400 K, and (d) mutated type BrCas12b at 400 K.

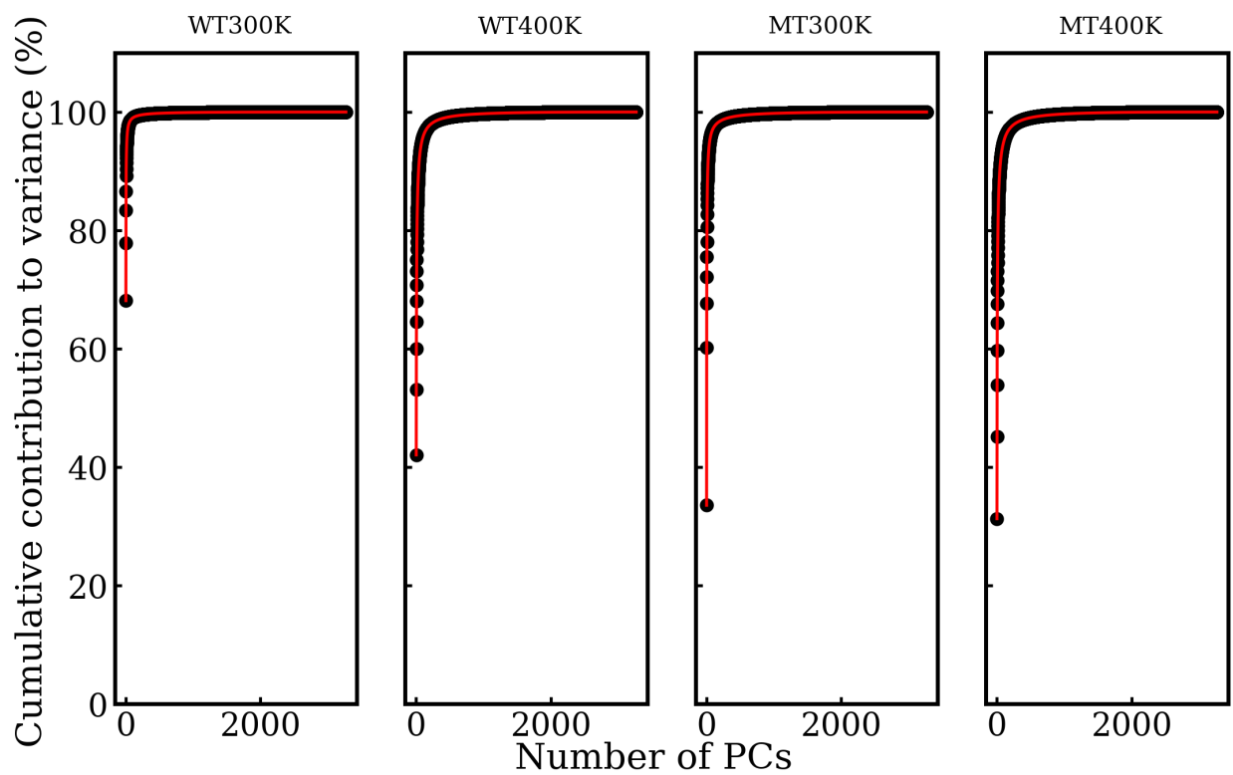

**Figure S10.** Cumulative contribution (%, y-axis) of all the principal components (PCs, x-axis) to the variance of the overall Cas12b motions calculated upon Principal Component Analysis (PCA) of WT and MT BrCas12b at 300 K and 400 K.